\documentclass[showpacs,aps,reprint,showpacs,onecolumn,tightenlines,superscriptaddress]{revtex4-1}
\usepackage[utf8]{inputenc}
\usepackage{amsmath, amsthm, amssymb, bm}
\usepackage{mathrsfs}
\usepackage{extarrows}
\usepackage{xcolor}
\usepackage{tabularx}
\usepackage{adjustbox}
\usepackage{multirow}
\usepackage{graphicx}
\usepackage[font=small,labelfont=bf]{caption}
\usepackage{braket}
\usepackage{pgfplots}
\usepackage[noline,ruled,linesnumbered]{algorithm2e}

\usepackage{caption}
\usepackage{qcircuit}
\usepackage{float}
\usepackage{hyperref}
\usepackage{subcaption}
\hypersetup{colorlinks=true, linkcolor=blue, citecolor=red, urlcolor=blue}

\newtheorem*{definition}{\indent Definition}

\newtheorem{lemma}{\indent Lemma}

\newtheorem{proposition}{Proposition}

\newcommand{\Cbb}{\mathbb{C}}
\newcommand{\Ebb}{\mathbb{E}}

\newcommand{\Ibb}{\mathbb{I}}

\newcommand{\Gcal}{\mathcal{G}}

\newcommand{\Qcal}{\mathcal{Q}}

\newcommand{\Tcal}{\mathcal{T}}

\newcommand{\poly}{\mathrm{poly}}

\newcommand{\im}{\mathrm{i}}

\newcommand{\var}{\mathrm{Var}}
\newcommand{\tr}[1]{\mathrm{Tr}\left( #1 \right)}

\newcommand{\abs}[1]{\left| #1 \right|}
\newcommand{\vabs}[1]{\left\| #1 \right\|}
\newcommand{\sbra}[1]{\left[ #1 \right]}
\newcommand{\pbra}[1]{\left( #1 \right)}
\newcommand{\cbra}[1]{\left\{ #1 \right\}}

\newcommand{\ketbra}[2]{|#1\rangle\langle #2|}

\newcommand{\TV}{\mathrm{TV}}

\begin{document}
\title{{Expectation value estimation with parametrized quantum circuits}}

\author{Bujiao Wu}
\email{bujiaowu@gmail.com}
\affiliation{International Quantum Academy, Shenzhen 518048, China}
\author{Lingyu Kong}
\affiliation{International Quantum Academy, Shenzhen 518048, China}
\affiliation{Shenzhen Institute for Quantum Science and Engineering,
Southern University of Science and Technology, Shenzhen 518055, China}

\author{Yuxuan Yan}
\affiliation{Center for Quantum Information, Institute for Interdisciplinary
Information Sciences, Tsinghua University, Beijing 100084, China}

\author{Fuchuan Wei}
\affiliation{Yau Mathematical Sciences Center and Department of Mathematics, Tsinghua University, Beijing 100084, China}

\author{Zhenhuan Liu}
\email{qubithuan@gmail.com}
\affiliation{Center for Quantum Information, Institute for Interdisciplinary
Information Sciences, Tsinghua University, Beijing 100084, China}

\begin{abstract}
Estimating properties of quantum states, such as fidelities, molecular energies, and correlation functions, is a fundamental task in quantum information science. 
{Due to the limitation of practical quantum devices, including limited circuit depth and connectivity, estimating even linear properties encounters high sample complexity.}
To address this inefficiency, we propose a framework that optimizes sample complexity for estimating the expectation value of any observable using a shallow parameterized quantum circuit. 
Within this framework, we introduce {two decomposition algorithms, a tensor network approach and a greedy projection approach} that decompose the target observable into a linear combination of multiple observables, each of which can be diagonalized with the shallow circuit. 
Using this decomposition, we then apply an importance sampling algorithm to estimate the expectation value of the target observable.
We numerically demonstrate the performance of our algorithm by estimating the {expectation values of some specific Hamiltonians and inner product of a Slater determinant with a pure state}, highlighting advantages compared to some conventional methods. 
Additionally, we derive the fundamental lower bound for the sample complexity required to estimate a target observable using a given shallow quantum circuit, thereby enhancing our understanding of the capabilities of shallow circuits in quantum learning tasks.

\end{abstract}
\maketitle

\section{Introduction}

{Learning the properties of an unknown quantum state is crucial across various fields, such as many-body physics,} quantum chemistry, and quantum information science~\cite{carleo2017solving,cao2019quantum,izmaylov2019revising,Eisert2020benchmarking,mcardle2020chemistry,Sharir2020Deep}. 
Among these properties, the linear properties $\tr{\rho H}$, where $H$ represents an observable, are of special importance since quantum mechanics is inherently linear. 
Therefore, various protocols have been developed to facilitate the estimation of linear properties, ranging from conventional tomography-based protocols~\cite{gross2010compress,haah2016tomo,Torlai2018neural_qst} and Pauli decomposition methods~\cite{yen2020measuring,verteletskyi2020measurement,Huang2021Efficient,hillmich2021decision,hadfield2021adaptive,miller2022hardwaretailored,Shlosberg2023Adaptive,wu2023overlapped,gresch2025guaranteed,zhang2023composite,yen2023deterministic} to quantum memory-assisted protocols~\cite{aaronson2018shadow,huang2022advantage}. 
Moreover, understanding the fundamental challenges in estimating linear properties can deepen our comprehension of quantum learning problems~\cite{huang2021bound,huang2022advantage,chen2022separation,chen2023adaptivity}.

Among all the estimation protocols, randomized measurement~\cite{Elben2023toolbox} has gradually become a prevailing toolbox due to its experimental accessibility~\cite{brydges2019probing,struchalin2021shadow,zhang2021shadow,hu2024demonstration,farias2024robust} and efficiency in estimating certain state properties~\cite{vermersch2019scrambling,elben2020mixed,zhou2020single,elben2020cross,cian2021chern,joshi2022otoc,liu2022detecting, huggins2022unbiasing}. Notably, Huang et al. introduced the classical shadow (CS) protocol~\cite{huang2020predicting}, which aims to provide a classical description of an unknown state $\rho$ and use it to calculate arbitrary state properties.
Conducted in a ``measure first, ask questions later'' manner~\cite{Elben2023toolbox}, the CS protocol does not require the information of observables during the quantum experiment and can efficiently estimate a large number of independent observables simultaneously.

Depending on the choice of unitary ensemble, the original classical shadow (CS) protocol can be categorized into two types: local CS and global CS, which sample unitaries from the product of single-qubit Clifford groups and the global $n$-qubit Clifford group, respectively. 
The local CS protocol is efficient for estimating $k$-local observables, where $k$ scales logarithmically with the number of qubits $n$. For general $k$-local observables, the sample complexity is $\mathcal{O}(4^k)$, and for $k$-fold tensor products of single-qubit observables, it improves to $\mathcal{O}(3^k)$. In contrast, the global CS protocol achieves $\mathcal{O}(1)$ sample complexity even for certain nonlocal observables, but requires implementing quantum circuits of logarithmic depth~\cite{schuster2024random}, which is still challenging for near-term devices. 
To make the CS protocol more suitable for noisy intermediate-scale quantum devices, a random shallow circuit-based CS protocol was developed~\cite{hu2024demonstration, hu2023scrambled,bertoni2024shallow,Akhtar2023scalableflexible,schuster2024random}. 
With a relatively low circuit depth, the shallow CS protocol outperforms the local CS protocols for certain nonlocal observables~\cite{ippoliti2023shadow, bertoni2024shallow, schuster2024random}. 
However, all these CS protocols are designed to be measurement-agnostic; they do not exploit prior knowledge about the target observables. As a result, they become suboptimal in scenarios where only a small number of observables need to be estimated. This limitation motivates approaches that incorporate knowledge of the observables to improve efficiency.

{In addition to the CS protocol, several methodologies have been developed to improve the efficiency of estimating the expectation values of specific observables with respect to certain quantum states~\cite{yen2020measuring, verteletskyi2020measurement, Huang2021Efficient, hillmich2021decision, hadfield2021adaptive, miller2022hardwaretailored, Shlosberg2023Adaptive, gresch2025guaranteed, zhang2023composite, wu2023overlapped, yen2023deterministic}.
}
Given the Pauli decomposition $H = \sum_{i=1}^m \alpha_i Q_i$, these methods aim to efficiently estimate the expectation values for all $\{Q_i\}_{i=1}^m$. 
This is achieved by performing Clifford circuits sampled from a specific set $\mathcal{S}$ with certain probabilities, followed by measurement in the computational basis and classical post-processing. 
The gate set $\mathcal{S}$ and these probabilities are carefully designed and vary depending on the specific algorithms.
While these methods improve performance compared to the original CS protocol by utilizing information about target observables, they do not consider the practical constraints of quantum circuits.
Firstly, implementing some target Clifford gates may require deep circuits~\cite{yen2020measuring,yen2023deterministic}. 
Additionally, since the observable decomposition is restricted to Pauli observables, these methods may still require high sample complexity even with deep circuits.

In this work, we propose a framework for estimating linear properties that unifies the logic of the methodologies mentioned earlier.
{Specifically, given an unknown quantum state and a variational circuit whose structure is determined by experimental settings, our framework aims to use the information of the target observable to optimize the usage of the parameterized circuit and achieve low sample complexity. 
}
Our approach is inspired by the observation that, given an observable $H$, if we can perform any unitary operation on the quantum state $\rho$, {$\mathcal{O}(\vabs{H}_2^2 \epsilon^{-2})$} copies of $\rho$ are sufficient to approximate $\tr{\rho H}$ with the additive error of $\epsilon$ and high success probability, where $\|\cdot\|_2$ denotes the spectral norm. 
This requirement arises from the eigen-decomposition of $H$, given by $H = U^\dagger \Lambda U$. 
Thus, we can rotate the quantum state using the unitary $U$ and measure it in the computational basis to estimate $\tr{\rho H}$.
In practical scenarios, where we are limited to $L$-depth parameterized unitaries $U_L(\theta)$ {which are tomography-complete}, we can decompose the target observable as $H \approx \sum_{k=1}^K U_L(\theta^{(k)})^\dagger \Lambda_k U_L(\theta^{(k)})$, with $\Lambda_k$ being real diagonal matrices. 
Each term in the summation is an observable that can be diagonalized with the parameterized circuit to efficiently estimate $\tr{\Lambda_k U_L(\theta^{(k)}) \rho U_L(\theta^{(k)})^\dagger }$. 

\begin{figure}[t]
\includegraphics[width=1.0\textwidth]{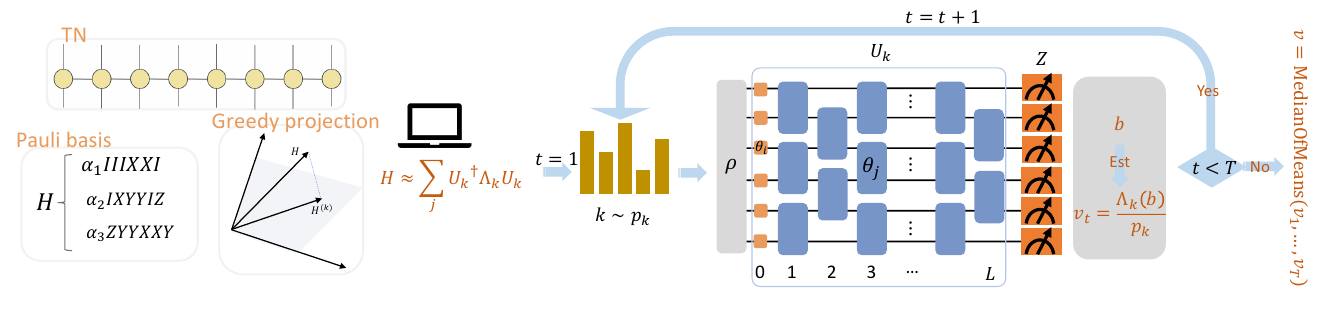}
 \caption{{Overview of the expectation value estimation with %shallow
 {parameterized} circuit. We utilize a classical algorithm (greedy projection or tensor network method) to generate a decomposition of $H \approx \sum_{k=1}^K U_L(\theta^{(k)})^\dagger \Lambda_k U_L(\theta^{(k)})$, followed by iteratively repeating $T$ rounds: sampling $k$ with probability $p_k\propto \vabs{\Lambda_k}_2$, performing $U_L(\theta^{(k)})$ on the prepared quantum state $\rho$, as illustrated by the $L$-depth circuit consisting of two-qubit parametrized gates. 
 Afterward, we measure in the computational basis to obtain a result $b$, which is then used to compute the estimator $ v $ by applying the Median-of-Means method.
 Here, $\Lambda_{k}(b)$ denotes the $b$-th diagonal entry of the diagonal matrix $\Lambda_k$ and $b$ ranges from $0$ to $2^n-1$.}}
    \label{fig:sketchAlg}
\end{figure}

Based on these observations, our framework can be divided into classical and quantum phases, as illustrated in Fig.~\ref{fig:sketchAlg}. 
In the classical phase, we use {a greedy projection or tensor network optimization} algorithm to decompose the target observable. 
In the quantum phase, we design an efficient method to estimate the expectation value of the target observable based on the given decomposition.
Many existing Pauli decomposition protocols, such as qubit-commuting measurement approaches~\cite{Huang2021Efficient,hadfield2021adaptive,wu2023overlapped} and commuting measurement approaches~\cite{yen2020measuring,yen2023deterministic}, fall within the scope of our framework. 
The required number of copies of $\rho$ is $\mathcal{O}\left(\frac{\|\bm{\Lambda}\|_1^2 \log(1/\delta)}{\epsilon^2}\right)$ with {additive error} $\epsilon$ and success probability $1-\delta$, where $\bm{\Lambda}$ is a $K$-dimensional vector with the $k$-th entry being $\bm{\Lambda}(k) = \|\Lambda_k\|_2$, and $\vabs{\bm\Lambda}_1$ denotes the sum of the absolute values of all the entries in $\bm\Lambda$.

To verify the convergence of our observable decomposition algorithm, we numerically evaluate the distance between the approximate observable $\sum_{k = 1}^K U_{L}(\theta^{(k)})^\dagger\Lambda_k U_{L}(\theta^{(k)})$ and the target observable across various circuit depths $L$ and the number of decomposition terms $K$. 
The numerical results indicate that for a fixed number of decomposition terms, this distance decreases as the depth of the ansatz circuits increases, which aligns with our intuition. 
Moreover, the Frobenius distance decays exponentially as the number of decomposition terms increases when fixing the circuit depth.
We further demonstrate the practical efficiency of our algorithms by applying them to estimate the ground energy of a sparse Hamiltonian and the inner product of two pure quantum states, comparing the results with conventional methods.

In addition to providing a practical measurement framework, we also aim to address the fundamental question: \emph{What is the limitation of a shallow quantum circuit in estimating an observable?} 
For a given variational quantum circuit $U_L(\theta)$, we establish a lower bound on the number of copies of the quantum state required to accurately estimate $\tr{\rho H}$. 
This bound is given by $\Omega\left(\frac{\tr{H_0^2}^2}{\epsilon^2 \delta(H_0) 4^n}\right)$, where $H_0$ represents the traceless part of $H$, and $\delta(H_0)$ is the square of the maximum expectation value of $H_0$ with respect to the quantum states prepared by applying $U_L(\theta)^\dagger$ to a product state.
We also analyze the tightness of this lower bound in specific cases.

\section{Expectation value estimation framework}
\label{sec:RMwithSA}

Our framework can be roughly divided into two phases.  
The classical phase relies on a specific decomposition algorithm to decompose the target observable into several terms. 
In the quantum phase, we estimate the expectation value based on the decomposition of the target observable and some probability distribution. 
We summarize the framework using Algorithm \ref{alg:frame_work}, in which $\vabs{\cdot}_2$ denotes the spectral norm, $\vabs{A}_2=\max_j \sigma_j$ with $\sigma_j$ being the singular value of matrix $A$, $\vabs{A}_F=\sqrt{\tr{AA^\dagger}}$ denotes the Frobenius norm of matrix $A$,  $\bm{\Lambda}$ represents a $K$-dimensional vector with the $k$-th entry being $ \bm\Lambda(k)= \vabs{\Lambda_k}_2$, and $\vabs{\bm\Lambda}_1$ denotes the sum of the absolute values of all the entries in $\bm\Lambda$.
{The $b$-th diagonal entry of $\Lambda_k$ is denoted by $\Lambda_k(b)$, where $b \in \{0,1\}^n$ is interpreted as an integer index in the computational basis.}

\SetNlSty{textbf}{}{\quad}
\IncMargin{1em}
\begin{algorithm}[htbp]
\SetKwInOut{Input}{input}
\SetKwInOut{Output}{output}
\Input{{Quantum state $\rho$, observable $H$, error $\epsilon_1$, the number of samplings $T$.}}
\Output{{Estimator $v$ of $\tr{\rho H}$} with error $\epsilon = \epsilon_1 + \epsilon_2$ and success probability $1 - \delta$.}
\emph{Generate unitaries $\cbra{U_k}_{k = 1}^K$ and real diagonal matrices $\cbra{\Lambda_k}_{k = 1}^K$, {where the $b$-th diagonal entry of $\Lambda_k$ is denoted by $\Lambda_k(b)$,} such that $\vabs{H - \sum_{k=1}^{K} U_k^\dagger \Lambda_k U_k}_2\leq \epsilon_1$\label{algstep:genAppH}\;}
\emph{{Set probabilities $p_k \propto \vabs{\Lambda_k}_2$}}\;
\For{$t\leftarrow 1$ \KwTo $T$}{
 \qquad \emph{Sample an index $k$ with probability $p_k$}\;   
 \qquad   \emph{Perform $U_k$ to the prepared state $\rho$, then measure in computational basis to obtain the $n$-bit measurement result $b_{t}$\label{alg_step:circuit_perform_1}\;
\qquad    Let the estimator $v_{t}=\frac{\Lambda_{k}(b_{t})}{p_k}$}\;
\label{alg_step:circuit_perform}
}
\Return{$v:=\operatorname{\mathbf{MedianOfMeans}}(v_1,\cdots,v_t)$}
\caption{Framework for Estimating Linear Properties of Quantum States.}
\label{alg:frame_work}
\end{algorithm}
\DecMargin{1em}

\begin{proposition}
Let $\hat{H}=\sum_{k=1}^K U_L\pbra{\theta^{(k)}}^\dagger \Lambda_k U_L\pbra{\theta^{(k)}}$ be an $\epsilon_1$ approximation of $H$ in spectral norm. Then $T=\mathcal{O}\left(\frac{\vabs{\bm{\Lambda}}_1^2\log(1/\delta)}{\epsilon_2^2}\right)$ copies of $\rho$ are required to estimate $\tr{\rho H}$ to additive error $\epsilon=\epsilon_1+\epsilon_2$ with success probability of at least $1-\delta$.
Here, $\bm{\Lambda}$ is a $K$-dimensional vector with the $k$-th  entry being $\vabs{\Lambda_k}_2$.
\label{prop:app_estimation}
\end{proposition}

\begin{proof}
Suppose there exists a set $\{\theta^{(k)}, \Lambda_k\}_{k = 1}^K$ such that $\vabs{H - \widehat{H}}_2 \leq \epsilon_1$ with $\widehat{H} = \sum_{k = 1}^K U_L(\theta^{(k)})^\dagger \Lambda_k U_L(\theta^{(k)})$, then
\begin{align}
\abs{\tr{\rho \widehat{H}} - \tr{\rho H}} \leq \vabs{\widehat{H} - H}_2 \leq \epsilon_1,
\label{eq:inequality_1}
\end{align}
where the first inequality holds since $\tr{\rho A} \leq \vabs{A}_2$ for any matrix $A$.

It remains to show that the estimation $v$ generated by Algorithm \ref{alg:frame_work} in combination with Algorithm \ref{alg:PD_LamDiagonalH} and important sampling is an unbiased estimation of $\tr{\rho \widehat{H}}$ and satisfies
\begin{align}
  \Pr\sbra{{\abs{v - \tr{\rho \widehat{H}}}\leq \epsilon_2}}\geq 1- \delta 
  \label{eq:inequality_2}
\end{align}
where $\delta$ represents the failure probability. 
Combining inequalities \eqref{eq:inequality_1} and \eqref{eq:inequality_2}, we obtain
\begin{align}
\abs{v - \tr{\rho H}}\leq \abs{v - \tr{\rho \widehat{H}}} + \abs{\tr{\rho \widehat{H}} - \tr{\rho H}}\leq \epsilon_1 + \epsilon_2,
\end{align}
with a success probability of $1 - \delta$. Let $v_t = \frac{\Lambda_k(b_t)}{p_k}$ be an estimation generated from a single sample, as shown in Step \eqref{alg_step:circuit_perform} of Algorithm \ref{alg:frame_work}. The expectation of $v_t$ can then be expressed as
\begin{align}
\Ebb\sbra{v_t} &= \sum_{k=1}^K\frac{1}{p_k} p_k \Ebb[\Lambda_k(b_{t})]\\
&= \sum_{k=1}^K \Pr[b] \Lambda_k(b)\\
&=\sum_{k=1}^K\tr{U_L(\theta^{(k)})\rho U_L(\theta^{(k)})^\dagger \Lambda_k}\\
&=\tr{\rho \widehat{H}}.
\end{align}
Then $\Ebb\sbra{v}=\Ebb\sbra{v_t} = \tr{\rho \widehat{H}}$.
Next, we will bound the variance of $v_t$ by noting that
\begin{align}
\var(v_t)&\leq \Ebb\sbra{v_t^2}\\
&=\sum_{k=1}^K \frac{p_k}{p_k^2} \tr{\rho U_L(\theta^{(k)})^\dagger \Lambda_k^2 U_L(\theta^{(k)})}\\
&\leq \sum_{k=1}^K \frac{1}{p_k}\vabs{\Lambda_k}_2^2\\
&=\sum_{k=1}^K \frac{\sum_{j=1}^K \vabs{\Lambda_j}_2}{\vabs{\Lambda_k}_2} \vabs{\Lambda_k}_2^2\\
&=\vabs{\bm \Lambda}_{1}^2
\end{align}
where $\bm\Lambda$ is a $K$-dimensional vector with the $k$-th  entry being $ \bm\Lambda(k)= \vabs{\Lambda_k}_2$, {and $\Lambda_k(b)$ denotes the $b$-th diagonal entry of the diagonal matrix $\Lambda_k$}.
According to the median-of-means method in Ref.~\cite{jerrum1986random}, with
$T=\mathcal{O}\left(\frac{\vabs{\bm \Lambda}_1^2 \log (1/\delta)}{\epsilon_2^2}\right)$ samplings, the estimation error for $v$ can be bounded to $\epsilon_1 + \epsilon_2$ in total with a success probability $1-\delta$.
\end{proof}

The central process of the classical stage is generating the decomposition $\{U_k, \Lambda_k\}_k$ of the target observable. 
In fact, many Pauli decomposition-based quantum measurement protocols offer methodologies for generating such decompositions.
Below, we will illustrate two of these methods: one based on qubit-commuting Pauli decomposition and the other on commuting Pauli decomposition.

\begin{itemize}
\item [(1)] \textbf{Qubit-commuting decomposition.}
We call Pauli operators $P$ and $Q$ qubit-commuting if, for any $l$-th qubit, $[P_l,Q_l]=0$. Since the $4^n$ Pauli operators form a basis for the space of $2^n \times 2^n$ Hermitian matrices, any observable $H$ can be expressed as $H = \sum_{i=1}^m \alpha_i Q_i$, where $\{\alpha_i\}_i$ are real coefficients and $\{Q_i\}_i$ are Pauli operators.
Since qubit-commuting Pauli observables can be estimated simultaneously, we can group the Pauli operators $Q_1, \ldots, Q_m$ into $K$ qubit-commuting (overlapping) sets $e_1, e_2, \ldots, e_K$, where each set contains Pauli operators that commute with each other. 
Various grouping methods have been proposed, such as the largest degree first algorithm~\cite{verteletskyi2020measurement} and the overlapping grouping algorithm~\cite{wu2023overlapped}.
Given a grouping algorithm, we can express the target observable as $H = \sum_{k=1}^K U_k^\dagger \Lambda_k U_k$, where each $U_k$ is the tensor product of local Clifford gates associated with the eigenstates of Pauli operators in the set $e_k$, and the diagonal matrix $\Lambda_k$ contains information about the eigenvalues of the Pauli operators in $e_k$. 
Further details on the explicit representations of $U_k$ and $\Lambda_k$ are provided in Appendix~\ref{app:qubit_commut_decomp}.
\item[(2)] \textbf{Commuting decomposition.} 
The qubit-commuting sets can be relaxed to the fully commuting sets $\cbra{g_k}_k$, where any two Pauli operators in $g_k$ commute with each other. 
Consequently, the target observable can be expressed as $H = \sum_{k=1}^K H_k$, with $H_k = \sum_{Q_j \in g_k} \alpha_j Q_j$. Similarly, there exists a set of Clifford gates $\{U_k\}_k$, which may require a polynomial-size quantum circuit to implement, and diagonal matrices $\{\Lambda_k\}_k$ such that $H_k = U_k^{\dagger} \Lambda_k U_k$~\cite{yen2020measuring} for all $k$. Details on how to generate $\{U_k, \Lambda_k\}_k$ are provided in Appendix~\ref{app:review_commut_meas}.
\end{itemize}
{We note that commuting decomposition may require polynomial-depth quantum circuits, making it unsuitable for near-term quantum devices. Nevertheless, its inclusion highlights the generality of our framework, which accommodates a wide range of measurement schemes. There are also alternative methods that leverage a more general basis. For instance, low-rank approximations are often applied to simplify electron-electron repulsion terms, reducing the required number of decomposition terms. Techniques such as density matrix embedding theory and density matrix renormalization group approximate the Hamiltonian by isolating essential degrees of freedom, thereby minimizing the number of terms necessary for an accurate representation~\cite{knizia2013density,Baiardi2020Density}.
In the following sections, we present two approaches for Hamiltonian approximation: (1) greedy projected decomposition (GPD) and (2) tensor network decomposition (TND).
}

The central objective of the quantum stage of our framework is to determine a probability distribution $\{p_k\}_k$ used to sample the estimated observables. 
In Pauli decomposition-based measurement protocols, this probability distribution corresponds to the one used for sampling different Pauli groups. 
Wu et al.~\cite{wu2023overlapped} demonstrated that all existing qubit-commuting protocols can be unified through two central processes: Pauli grouping and assigning each group a probability, which fits precisely within our framework.
Similarly, we show in Appendix~\ref{app:review_commut_meas} that existing commuting measurement schemes~\cite{yen2020measuring, yen2023deterministic} also fit within this framework.
It is important to note that Pauli decomposition-based protocols are limited to Clifford unitary evolution, which cannot achieve the optimal performance for a general observable even with arbitrary Clifford unitaries. 
Therefore, our goal is to design new algorithms that are compatible with general quantum circuits.

\section{{Greedy projection decomposition of the Hamiltonian}}
\label{subsec:framework}

Based on the introduced framework, we design specific algorithms to generate the decomposition of the target observable and estimate the expectation value, as illustrated in Fig.~\ref{fig:sketchAlg}.
We consider a practical scenario, where the depth and structure of the quantum circuit are given according to the capabilities of the available quantum devices and can go beyond Clifford circuits.
Without loss of generality, we hereafter take a typical circuit structure: an $L$-depth parameterized quantum circuit $U_L(\theta)$ with the brick-like arrangement of 2-qubit parametrized gates, to demonstrate our algorithms.
The 0-layer of the ansatz circuit comprises a layer of parametrized single-qubit gates. 
This parameterized circuit is tomography-complete for any $L \geq 0$ as it can estimate all Pauli observables, even when $L = 0$. 
Our conclusions can be easily generalized to other circuit structures.

We propose a greedy algorithm to decompose the target observable $H$ using the $L$-depth parameterized circuit $U_L(\theta)$. 
In each step, we aim to find an observable, $U_L(\theta)^\dagger\Lambda U_L(\theta)$, that approximates $H$. 
This involves finding a unitary $U_L(\theta)$ and a real diagonal matrix $\Lambda$ that minimize the distance between $U_L(\theta) H U_L(\theta)^\dagger$ and $\Lambda$.
When using the Frobenius norm to quantify the distance, the optimal choice of $\Lambda$ is $\Lambda = \mathrm{diag}[U_L(\theta) H U_L(\theta)^\dagger]$, where $\mathrm{diag}(X)$ denotes a diagonal matrix whose elements are the diagonal elements of $X$. 
As a comparison, the distance based on the 2-norm brings us an additional optimization step to determine the optimal $\Lambda$.
This iterative process updates the Hamiltonian $H$ as outlined in Algorithm \ref{alg:PD_LamDiagonalH}.

\begin{algorithm}[htbp]
\SetKwInOut{Input}{input}
\SetKwInOut{Output}{output}
\Input{Observable $H$, $L$-depth parameterized quantum circuit $U_L(\theta)$, accuracy $\epsilon$.}
\Output{Set $\cbra{\theta^{(k)},\Lambda_k}_k$ such that $\vabs{H - \sum_{k} U_L(\theta_k)^\dagger \Lambda_k U_L(\theta_k)}_2\leq \epsilon$.}
\emph{$H^{(0)} = H$}\;
\emph{$k = 0$}\;
\While{$\vabs{H^{(k)} }_2\geq \epsilon $}{
\qquad \emph{$\theta^{(k)} = \arg\min_{\theta}\vabs{U_L \pbra{\theta}H^{(k)}U_L^\dagger \pbra{\theta} - \mathrm{diag}\left[U_L \pbra{\theta}H^{(k)}U_L^\dagger \pbra{\theta}\right] }_F$\label{alg_step:ite_classical}}\;

\qquad \emph{$\Lambda_k = \mathrm{diag}\left[U_L \pbra{\theta^{(k)}}H^{(k)}U_L^\dagger \pbra{\theta^{(k)}}\right]$}\;

\qquad \emph{$H^{(k+1)}=H^{(k)} - U_L^\dagger\pbra{\theta^{(k)}}\Lambda_k U_L \pbra{\theta^{(k)}}$\;
\qquad $k = k + 1$}\;
}
\Return{$\cbra{\theta^{(k)},\Lambda_k}$}
\caption{Greedy Projected Decomposition (GPD) Algorithm.}
\label{alg:PD_LamDiagonalH}
\end{algorithm}

Using the decomposition generated by Algorithm~\ref{alg:PD_LamDiagonalH}, we claim that Algorithm~\ref{alg:frame_work} can produce an estimator of $\tr{\rho H}$ with bounded error.

\begin{proposition}
Given that $L$-depth unitary set $\cbra{U_L(\theta)}$ is tomography-complete and integer $L\geq 0, \epsilon\in(0,1)$, then any Hermitian matrix $H$ can be approximated as $\hat{H } = \sum_{k=1}^K U_L\pbra{\theta^{(k)}}^\dagger \Lambda_k U_L\pbra{\theta^{(k)}}$ such that $\vabs{\hat{H} - H}\leq \epsilon$, for some integer $K$ and diagonal matrix $\Lambda_k$.
\label{prop:app_hermitian}
\end{proposition}

\begin{proof}
Below, we will prove that Algorithm \ref{alg:PD_LamDiagonalH} can be conclusively terminated within a finite number of steps, denoted as $K$.
Considering one step of parameterized searching, the goal is to minimize the distance $\vabs{H-U_L\pbra{\theta}^\dagger \Lambda U_L\pbra{\theta}}_F = \vabs{U_L\pbra{\theta}H U_L\pbra{\theta}^\dagger - \Lambda}_F$.
Since $\Lambda$ is a real diagonal matrix, the optimal choice is $\Lambda = \mathrm{diag}\left[U_L\pbra{\theta}H U_L\pbra{\theta}^\dagger\right]$. This choice is optimal because the Frobenius norm quantifies the sum of the squares of all the matrix elements, and choosing $\Lambda$ in this way eliminates all the diagonal terms of $U_L\pbra{\theta}H U_L\pbra{\theta}^\dagger$.
Furthermore, with this choice of $\Lambda$, one can prove that $\vabs{H-U_L\pbra{\theta}^\dagger \Lambda U_L\pbra{\theta}}_F = \vabs{U_L\pbra{\theta}H U_L\pbra{\theta}^\dagger - \Lambda}_F \le \vabs{U_L\pbra{\theta}H U_L\pbra{\theta}^\dagger}_F = \vabs{H}_F$.
This means that when the optimization algorithm functions ideally, a better approximation of $H$ can always be achieved by increasing the iteration steps.

Furthermore, we can prove that it is always possible to find a $\theta$ such that $\vabs{H - U_L\pbra{\theta}^\dagger \Lambda U_L\pbra{\theta}}_F < \vabs{H}_F$, provided our parameterized circuit is tomography-complete. 
If this is not the case, then $\mathrm{diag}[U_L\pbra{\theta}H U_L\pbra{\theta}^\dagger] = 0$ must hold for all choices of $\theta$. If $H \neq 0$, this implies the existence of a non-zero observable that cannot be estimated using our parameterized circuits, which contradicts our assumption.
{We can further prove that if $\theta$ is the optimal value we can find, then $\vabs{H}_F - \vabs{H - U_L\pbra{\theta}^\dagger \Lambda U_L\pbra{\theta}}_F \ge \frac{1}{2\times4^n} \vabs{H}_F$. This is because we assume that $U_{L=0}(\theta)$ can realize the tensor product of arbitrary single-qubit gates, thereby enabling the estimation of arbitrary Pauli observables.
Given the Pauli decomposition of the observable $H = \sum_i \alpha_i P_i$, one can at least choose $U_L\pbra{\theta}^\dagger \Lambda U_L\pbra{\theta} = \alpha_j P_j$, where $\alpha_j$ is the largest coefficient. 
In this case, $\vabs{H}_F - \vabs{H - \alpha_j P_j}_F \ge \frac{1}{2\times4^n} \vabs{H}_F$.
Considering that the greedy algorithm can achieve better performance than simply picking the largest Pauli term, we can conclude that our algorithm can terminate in a finite number of steps.}
\end{proof}

{In Step 4 of Algorithm \ref{alg:PD_LamDiagonalH}, the classical processing time required is $ \mathcal{O}(\text{poly}(n) \cdot 2^{\omega n}) $, where $ \omega = 2.371552 \dots $ represents the matrix multiplication exponent \cite{williamsBounds2024}. This complexity is an improvement over the direct eigen-decomposition time of $ \mathcal{O}(2^{3n}) $ \cite{golub2013matrix}. Additionally, the matrix multiplication process can be further optimized with a quantum-inspired algorithm \cite{tang2019quantum}, allowing for a $\text{poly}(n)$-rank approximation to be performed in $\text{poly}(n)$ time.}

Here, the total error $\epsilon$ depends on both the approximation error of $H$ and the estimation error of $\tr{\rho \widehat{H}}$, where $\widehat{H}=\sum_{k=1}^K U_L\left(\theta^{(k)}\right)^\dagger \Lambda_k U_L\left(\theta^{(k)}\right)$. 
To minimize the resource consumption, we aim to minimize the value of $K$ while keeping a small approximation error.
{Due to the greedy nature of Algorithm \ref{alg:PD_LamDiagonalH}, theoretical analysis of the number of decomposed terms $K $ is challenging. However, numerical results indicate that the number of decomposed terms generated by our algorithm decreases as the depth $L $ increases.}
In the numerical section, we demonstrate that our algorithm, based on parameterized circuits, requires significantly fewer terms ($K$) to approximate a sparse and dense Hamiltonian within a given error threshold, compared to the number of terms needed in the standard qubit-wise Pauli decomposition approach.

\section{{Tensor network decomposition of the Hamiltonian}}

\begin{figure}
    \centering
    \includegraphics[width=1.0\linewidth]{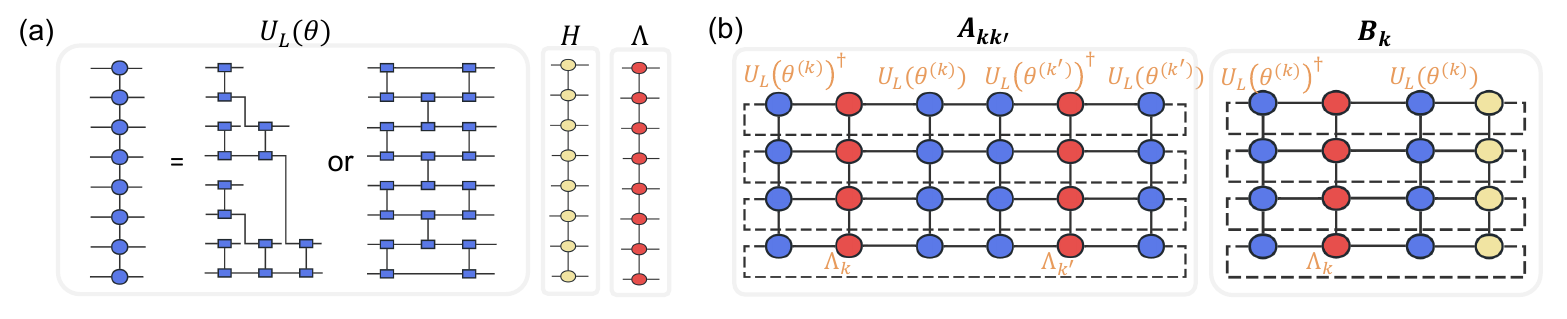}
    \caption{{Illustration of tensor network decomposition for a given Hamiltonian.
(a) Tensor network representation of the parametrized unitary $U(\theta_k)$, the diagonal matrix $\Lambda_k$, and the Hamiltonian $H$.
(b) Tensor network representation of $A_{kk'}$ and $B_k$, which are associated with the loss function calculations. } }
    \label{fig:Tensor Network Variables}
\end{figure}

Although the greedy decomposition method works generally for all parameterized circuits, it can be further optimized in some practical cases.
When the depth of the parameterized circuit is low, a tensor network–based approach can be employed to approximate the Hamiltonian as $\hat{H} = \sum_{k = 1}^{K} U_L(\theta_k)^\dagger \Lambda_k U_L(\theta_k)$.
Each $U_L(\theta_k)$ corresponds to a unitary tensor network of depth $L$, such as a unitary matrix product operator or a unitary tree tensor network, constructed following~\cite{Verstraete2004matrix,Shi2006classical,Huang2023Tensor}, as illustrated in~Fig.~\ref{fig:Tensor Network Variables}(a). The bond dimension is $O(nr_U^{})$
The diagonal matrices $\Lambda_k$ are represented as matrix product operators (MPOs), with component-wise form given by $\Lambda_{i_1,\ldots,i_n,l_1,\ldots,l_n} = \bigotimes_{j=1}^n \Lambda_{j,i_j,l_j,u_{j-1},u_j}$, where $u_j \in [\beta]$ for $1 \leq j \leq n-1$, and the boundary conditions are $u_0 = 1$, $u_n = 1$. 
The indices $i_j, l_j \in \cbra{0,1}$ denote the input and output basis states, respectively, as depicted in Fig.~\ref{fig:Tensor Network Variables}(a). The target Hamiltonian $H$ is also expressed in MPO form.

To approximate the Hamiltonian $H$, we need to minimize the following loss function:
\begin{align}
\mathcal{L} &=
\| H - \sum_{k} U_L(\theta_k)^\dagger \Lambda_k U_L(\theta_k) \|_F^2\\
&= \tr{H^2} + \sum_{k,k'}\tr{ U_L(\theta_k)^\dagger \Lambda_k U_L(\theta_k)^\dagger U_L(\theta_{k'}) \Lambda_{k'} U_L(\theta_{k'}) } - 2 \, \text{Re} \sum_k \tr{ U_L(\theta_k)^\dagger \Lambda_k U_L(\theta_k)H } \\
&=: \sum_{k,k'}A_{kk'} - 2 \, \text{Re} \sum_k B_k + c,
\label{eq:loss_function_tn}
\end{align}
where we define $A_{kk'} := \tr{ U_L(\theta_k)^\dagger \Lambda_k U_L(\theta_k)^\dagger U_L(\theta_{k'}) \Lambda_{k'} U_L(\theta_{k'}) }, B_k  := \tr{ U_L(\theta_k)^\dagger \Lambda_k U_L(\theta_k)H },$ and $c := \tr{H^2}$ is a constant independent of the parameters $\cbra{\theta^{(k)}, \Lambda_k}$.
The loss function can also be formulated as a tensor network contraction, where the first term corresponds to the tensor $A_{kk'}$, and the second term to the tensor $B_k$, both illustrated in Fig.~\ref{fig:Tensor Network Variables} (b). The quantities $A_{kk'}$ and $B_k$ can be efficiently computed by contracting the corresponding tensor networks depicted in the two diagrams, where the diagrams are connected in a head-to-tail fashion. The overall contraction cost remains polynomial in system size provided that $K = \poly(n)$, the circuit depth $L= O\pbra{\log (n)}$, and both $\Lambda_k$ and $H$ admit efficient MPO representations.

Once the contraction is performed, the loss function $\mathcal{L}$ can be minimized using standard gradient descent optimization techniques. By keeping the parameters of all other terms fixed, we can iteratively optimize the parameters $\cbra{\theta^{(k)}, \Lambda_k}$ for each term $k$, thereby constructing an approximation of the target Hamiltonian.

\section{Numerical results}
\label{sec:numerics}

Estimating expectation values for Hamiltonians is one of the most crucial tasks in state learning, and numerous algorithms have been developed for this purpose ~\cite{huang2020predicting,Huang2021Efficient,wu2023overlapped,zhang2023composite}. In our numerical experiments, each two-qubit parameterized gate consists of an iSWAP gate followed by a tensor product of two arbitrary single-qubit gates.

We assess the performance of the GPD algorithm through three representative examples. The first and second examples focus on estimating the expectation values of randomly generated sparse and dense Hamiltonians, respectively. The third example involves estimating the inner product between a Slater determinant $\ket{\phi_\tau}$ and a given pure state $\ket{\psi}$, denoted as $\braket{\phi_\tau|\psi}$. The convergence behavior for all three cases is provided in Appendix~\ref{app:supp_numerics}.

We demonstrate the effectiveness of the tensor network decomposition algorithm by showing its significant performance advantages over existing qubit-wise commuting algorithms and the classical shadow method.

The numerical simulations were conducted mainly using MindSpore Quantum~\cite{mq_2021}.

\section{Ground energy estimation}
Sparse Hamiltonians play important roles in quantum computation, chemistry, and simulation~\cite{berry2007efficient,Harrow2009quantum,changlani2015density}. {Here, we demonstrate the performance of the GPD algorithm with parameterized quantum circuits by estimating the ground state energy of a randomly generated sparse Hamiltonian. We compare our results with state-of-the-art qubit-commuting methods~\cite{Huang2021Efficient, hadfield2021adaptive, wu2023overlapped, zhang2023composite, gresch2025guaranteed} and the standard global CS algorithm~\cite{huang2020predicting}.}

\begin{table}[h]
    \centering
        \caption{{Estimation errors for the ground-state energy of a sparse Hamiltonian across different algorithms. The system consists of $n = 8$ qubits, and the GPD algorithm uses a quantum parameterized circuit with depth $L = 4$ and employs $K = 80$ decomposition terms.}}
\begin{tabular}{rcccccccc}
\hline
\textbf{Copies} & \textbf{CS~\cite{huang2020predicting}} & \textbf{Derand~\cite{Huang2021Efficient}} & \textbf{C-LBCS~\cite{zhang2023composite}}
&
\textbf{SG~\cite{gresch2025guaranteed}} & \textbf{Random~\cite{huang2020predicting}} & \textbf{Adaptive~\cite{hadfield2021adaptive}} & \textbf{OGM~\cite{wu2023overlapped}} & \textbf{GPD}\\
\hline
45      & 1.095 & 0.616 & 1.140 & 1.247& 1.415 & 1.052& 0.738& \textbf{0.674}\\
160     & 0.584 & 0.556 & 0.708 & 0.628& 1.328 & 0.831& 0.522& \textbf{0.312}\\
572     & 0.260 & 0.361 & 0.357 & 0.273& 0.637 & 0.505& 0.416& \textbf{0.203}\\
2038    & 0.145 & 0.232 & 0.184 & 0.171& 0.275 & 0.248& 0.191& \textbf{0.104}\\
7256    & 0.066 & 0.137 & 0.142 & 0.082& 0.161 & 0.115& 0.122& \textbf{0.046}\\
25848   & 0.050 & 0.069 & 0.059 & 0.053& 0.084 & 0.065& 0.097& \textbf{0.030}\\
\hline
\end{tabular}
    \label{tab:sparseH_error}
\end{table}

The numerical results are shown in Table~\ref{tab:sparseH_error}.
Here, we randomly generate an 8-qubit sparse Hamiltonian $H$ with $64=8^2$ non-zero matrix elements and the ground energy of approximately $-1.536$. 
In our implementation, we set the number of decomposition terms to $K = 80$ and the depth of the quantum ansatz to $L = 4$, which is significantly smaller than the total number of Pauli terms, $4^8$. Detailed results on how the approximation error between $\hat{H}$ and $H$ depends on the number of decomposition terms—using unitary ansätze with depths ranging from ${0, 1, 2, 3}$—are provided in Appendix~\ref{app:supp_numerics}. These results demonstrate that the error decays exponentially with the number of decomposition terms.
We further compare our method with existing algorithms, showing that the estimation error achieved by our approach is substantially lower than those of prior methods, illustrating the advantage of our algorithm in estimating the expectation value of sparse Hamiltonians. Specifically, ``CS'' refers to the global classical shadows protocol~\cite{huang2020predicting}, which applies a global Clifford operation before measurement, while ``Random'' denotes the local classical shadows protocol introduced in the same reference.

\begin{table}[h]
    \centering
    \caption{{Estimation errors for the expectation value of a randomly generated dense Hamiltonian across different algorithms. The number of qubits is fixed at $n = 4$, and the GPD algorithm uses a quantum parameterized circuit with depth $L = 4$ and employs $K = 20$ decomposition terms.}}
\begin{tabular}{rcccccccc}
\hline
\textbf{Copies} & 	\textbf{CS~\cite{huang2020predicting}}  & 	\textbf{Derand~\cite{Huang2021Efficient}} & 	\textbf{C-LBCS~\cite{zhang2023composite}}&	\textbf{SG~\cite{gresch2025guaranteed}} & 	\textbf{Random~\cite{huang2020predicting}} & 	\textbf{Adaptive~\cite{hadfield2021adaptive}} & 	\textbf{OGM~\cite{wu2023overlapped}} & 	\textbf{GPD}\\
\hline
45      & 1.702 & 1.197 & 1.062 & 1.737& 1.586 & 1.876&  1.152& \textbf{0.922}\\
160     & 0.728 & 0.536 & 0.742 & 0.881& 1.150 & 1.261&  0.556& \textbf{0.503}\\
572     & 0.414 & 0.296 & 0.381 & 0.539& 0.707 & 0.593&  0.293& \textbf{0.253}\\
2038    & 0.243 & 0.190 & 0.206 & 0.262& 0.353 & 0.313&  0.157& \textbf{0.147}\\
7259    & 0.103 & 0.081 & 0.095 & 0.124& 0.228 & 0.159&  0.085& \textbf{0.061}\\
25848   & 0.077 & 0.064 & 0.243 & 0.059& 0.096 & 0.079&  0.053& \textbf{0.046}\\
\hline
\end{tabular}
    \label{tab:denseH}
\end{table}

 %\emph{.}--
 \subsection{Expectation value estimation for a dense Hamiltonian}
 To evaluate the effectiveness of the GPD algorithm with parameterized circuits in a general setting, we consider the task of estimating the expectation value of a randomly generated dense Hamiltonian. Specifically, we construct a random matrix $A$ with entries independently sampled from the uniform distribution over $[0,1]$, and define the Hamiltonian as $H := (A + A^T)/2$ to ensure Hermiticity. The input quantum state is also randomly generated with amplitudes drawn uniformly from $[0,1]$. The decomposition is performed using $K = 20$ terms with a parameterized quantum ansatz of depth $L = 4$. We compare the estimation performance of our method against existing algorithms as the number of copies of the quantum state increases. As shown in Table~\ref{tab:denseH}, the GPD algorithm consistently yields significantly lower estimation errors, demonstrating its effectiveness in general-purpose Hamiltonian estimation tasks. We also report the average estimation error with the number of samplings in the expectation values for several randomly selected Hamiltonians (both sparse and dense) in Appendix~\ref{app:supp_numerics}.

\begin{table}[h]
    \centering
        \caption{{Estimation errors for the inner product between a 3-qubit Slater determinant and a 3-qubit pure state across different algorithms. The GPD algorithm uses a quantum parameterized circuit with depth $L = 4$ and employs $K = 20$ decomposition terms. }}
\begin{tabular}{rccccccccc}
\hline
\textbf{Copies} & \textbf{CS~\cite{huang2020predicting}} & \textbf{FCS~\cite{wan2023matchgate}} & \textbf{Derand~\cite{Huang2021Efficient}} & \textbf{C-LBCS~\cite{zhang2023composite}}&\textbf{SG~\cite{gresch2025guaranteed}} & \textbf{Random~\cite{huang2020predicting}} & \textbf{Adaptive~\cite{hadfield2021adaptive}} & \textbf{OGM~\cite{wu2023overlapped}} & \textbf{GPD}\\
\hline
45      & 0.366 & 0.273 & 0.295 & 0.245 & 0.309& 0.355 & 0.311& 0.278& \textbf{0.222}\\
160     & 0.195 & 0.160 & 0.150 & 0.200 & 0.165& 0.260 & 0.176& 0.159& \textbf{0.112}\\
572     & 0.108 & 0.082 & 0.072 & 0.128 & 0.082& 0.126 & 0.080& 0.078& \textbf{0.056}\\
2038    & 0.056 & 0.041 & 0.041 & 0.067 & 0.043& 0.072 & 0.043& 0.046& \textbf{0.030}\\
7256    & 0.027 & 0.025 & 0.023 & 0.039 & 0.024& 0.044 & 0.024& 0.023& \textbf{0.015}\\
25848   & 0.015 & 0.014 & 0.017 & 0.086 & 0.012& 0.023 & 0.012& 0.012& \textbf{0.009}\\
\hline
\end{tabular}
    \label{tab:slater}
\end{table}

\emph{Inner product estimation}--GPD algorithm with parameterized circuits can be easily generalized to estimate non-Hermitian operators by allowing $\{\Lambda_k\}_k$ to be complex-valued diagonal matrices.
Hereby, we apply our algorithm to estimate the expectation value of a non-Hermitian operator $O = \ket{1}\ket{\phi_{\tau}}\bra{0^{n+1}}$. 
This operator enables the calculation of the inner product between a Slater determinant and a pure quantum state, where $\ket{\phi_{\tau}}$ represents a $\tau$-Slater determinant. This has applications in algorithms such as the QC-AFQMC algorithm \cite{huggins2022unbiasing}. 
Here, $\tau$-Slater determinant~\cite{wan2023matchgate} is defined as $\ket{\phi_\tau} = \tilde{b}_1^\dagger \tilde{b}_2^\dagger\ldots \tilde{b}_\tau^\dagger\ket{0^n}$, where $\tilde{b}_{k} = \sum_{j=1}^n U_{kj} b_j$, $U$ is an $n\times n$ unitary operator, and $\{b_j\}_j$ are annihilation operators.

Provided with the circuit information for preparing $\ket{\psi}$, the inner product of a $\tau$-Slater determinant $\ket{\phi_\tau}$ and a pure state $\ket{\psi}$ can be estimated with our algorithms.
We start by preparing the state $\ket{\psi'}=\frac{\ket{0^{n+1}} + \ket{1}\ket{\psi}}{\sqrt{2}}$ using an ancillary qubit and use it as the input state for our algorithms~\cite{wu2023overlapped}.
We then choose the operator $O = \ket{1}\ket{\phi_\tau}\bra{0^{n+1}}$, and hence $\braket{\psi|\phi_\tau} = 2\tr{\ketbra{\psi'}{\psi'} O}$.
To demonstrate our algorithms, we set $\tau = 1$, $n = 3$, and use a $3 \times 3$ random unitary $U$. 
We randomly generate a 3-qubit quantum state $\ket{\psi}$ with $\braket{\psi|\phi_\tau} \approx -0.0446+0.0552\im$. 
We postpone the distance of the estimated $\hat{O}$ with the ideal $O$ into Appendix~\ref{app:supp_numerics}. 
We give the estimation error for $\braket{\psi|\phi_\tau}$ of our algorithm and existing algorithms in Table~\ref{tab:slater}. 
For comparison, we also estimate the errors using existing state-of-the-art qubit-wise commuting measurements~\cite{Huang2021Efficient,zhang2023composite,gresch2025guaranteed,hadfield2021adaptive,wu2023overlapped}, global/local CS algorithm~\cite{huang2020predicting} and fermionic CS algorithm~\cite{wan2023matchgate}. 
The numerical result illustrates the advantage of our algorithm in estimating the inner product of a Slater determinant and a quantum pure state.
Detailed explanations of the error estimation for these existing algorithms are provided in Appendix~\ref{app:pre_inputstate}.

%\emph{}--
\subsection{Expectation value calculation with TND}
Here we consider the estimation of the expectation value for a Hamiltonian that admits a matrix product operator (MPO) representation. Specifically, we study the Hamiltonian
\begin{align}
  W = h_0X_1\ldots X_n + \sum_{j=1}^{n-1} h_i X_1\ldots X_{j-1} Z_j Z_{j+1}.
\end{align}
We express $W$ in MPO form as $W = J W^{[1]} W^{[2]} \ldots W^{[n]}$, where the boundary tensors are given by
$W^{[1]} = \begin{pmatrix} \mathbb{I} & -Z_1 & g X_1 \end{pmatrix}$ and $W^{[n]} = \begin{pmatrix} g X_n & -Z_n & \mathbb{I}_n \end{pmatrix}^T$. The bulk tensors for $i \in \cbra{2, \ldots, n-1}$ are defined as
\begin{align}
    W^{[i]} = \begin{pmatrix}
        \Ibb_i & -Z_i & gX_i\\
        0 & 0 & Z_i\\
        0 & 0 & \Ibb_i
    \end{pmatrix}.
\end{align}

 In our numerical experiments, we set $J = 1$ and $g = 1$. The optimal parameters for the unitary ansatz are obtained by minimizing the loss function defined in Eq.~\eqref{eq:loss_function_tn}.

To evaluate the effectiveness of the TND method, we generate a random 8-qubit quantum state $\rho$ and compare the estimation error with that of existing algorithms, as shown in Table \ref{tab:TN_numerics}. In our algorithm, the Hamiltonian is decomposed into three terms, and the depth of the variational quantum circuit is set to $1$, with the circuit structure matching that of the GPD algorithm. The results demonstrate that our approach outperforms traditional Pauli-based measurement schemes. We give the approximation error of the Hamiltonian $W$ in Appendix~\ref{app:supp_numerics}. To demonstrate the robustness of the evaluation with respect to different quantum input states, we also present the average errors for five randomly selected input states in Appendix~\ref{app:supp_numerics}, further supporting the advantages of our TND algorithm with parameterized quantum circuits.

\begin{table}[h]
    \centering
        \caption{{Estimation errors for the low bond dimension Hamiltonian across different algorithms. The system consists of $n=8$ qubits. The TND algorithm uses a quantum parameterized circuit with depth $L=1$ and employs $K = 3$ decomposition terms.}}
\begin{tabular}{rcccccccc}
\hline
\textbf{Nsteps} & \textbf{CS~\cite{huang2020predicting}} & \textbf{Derand~\cite{Huang2021Efficient}} & \textbf{C-LBCS~\cite{zhang2023composite}}&\textbf{SG~\cite{gresch2025guaranteed}} & \textbf{Random~\cite{huang2020predicting}} & \textbf{Adaptive~\cite{hadfield2021adaptive}} & \textbf{OGM~\cite{wu2023overlapped}} & \textbf{TND}\\
\hline
1000    & 4.616 & 0.638 & 0.535 & 0.369& 5.892 & 0.445& 0.448& \textbf{0.213}\\
3000    & 3.029 & 0.700 & 0.250 & 0.227& 5.932 & 0.217& 0.250& \textbf{0.103}\\
9000    & 1.859 & 0.605 & 0.156 & 0.117& 6.165 & 0.121& 0.178& \textbf{0.079}\\
18000   & 2.292 & 0.615 & 0.105 & 0.108& 4.475 & 0.080& 0.094& \textbf{0.050}\\
\hline
\end{tabular}
    \label{tab:TN_numerics}
\end{table}

\section{Lower bound for sample complexity}
\label{sec:lowerbound}
This work aims to harness the power of 
parameterized
circuits for observable estimation. 
The numerical results presented in the previous section show that performance is closely tied to circuit depth.
This naturally raises the question of whether this dependence stems from our algorithms or if there are inherent limitations to the performance of shallow circuits. In this section, we aim to conduct a protocol-independent analysis of shallow circuit performance. Such an analysis will not only help us evaluate our algorithms but also deepen our understanding of the capabilities of shallow quantum circuits in state learning tasks.

Here, we establish a fundamental lower bound on the number of copies of the quantum state $\rho$ required to accurately estimate the value of $\tr{\rho H}$. This lower bound holds for all single-copy adaptive strategies that use the given parameterized circuit $U_L(\theta)$ without ancillary systems. Our approach builds on the techniques introduced by Chen et al.~\cite{Chen2021exponential}.
While their work does not cover scenarios where a limited quantum ansatz is utilized.

\begin{proposition}[Sample Complexity Lower Bound]\label{proposition:SampleComplexity_LB}
Suppose $H\in \Cbb^{2^n\times 2^n}$ is a Hermitian operator we aim to estimate.
Denote $\mathcal{S}:=\{U_L(\theta)^\dagger\ket{b}\mid \forall\theta ,b\in\{0,1\}^n\}$.
Then for any single-copy adaptive strategy relying solely on the parameterized circuit $U_L(\theta)$ executed on $\rho$, the sample complexity $T$ for measuring $\tr{\rho H}$ to accuracy $\epsilon$, for arbitrary unknown state $\rho$, with success probability of at least $2/3$, is lower bounded by
\begin{equation}
T=\Omega\left(\frac{\tr{H_0^2}^2}{\epsilon^2\delta(H_0)4^n}\right),
\label{eq:LowerboundT}
\end{equation}
where $H_0=H-\frac{\tr{H}}{2^n}\mathbb{I}_{2^n}$ denotes the traceless part of $H$ and $\delta(H_0):=\underset{\ket{\psi} \in \mathcal{S}}{\sup}\bra{\psi}H_0\ket{\psi}^2$.
\label{pro:lowerboundAnsatz}
\end{proposition}

{A learning algorithm that uses $T $ copies of measurements to learn a quantum state $\rho $ without quantum memory can be modeled as a rooted tree $\mathcal{T} $ with depth $T $. The central idea behind proving the proposition is to demonstrate that for a Hamiltonian $H $, there exist two quantum state sets $\{ \rho_x \}_x $ and $\left\{ \frac{\mathbb{I}}{2^n} \right\} $, such that distinguishing between the expected values $\tr{\rho_x H} $ and $\tr{ \frac{H}{2^n} } $ using the quantum learning algorithm requires at least $T $ samples. This process is carried out with the condition that $L $-depth quantum circuits are applied before measurement in the computational basis. The problem then reduces to bounding the total variation distance between two distributions generated at the leaves of the rooted tree $\mathcal{T} $, which quantifies the distinguishability of the distributions produced by the quantum algorithm with $T $ copies of the quantum states. Further details of the proof appear in Appendix~\ref{app:LowerBound}. This bound provides the desired performance guarantee for the algorithm in terms of the required number of samples.
}

Here, the dependence of this lower bound on the capability of the circuit is mainly reflected on $\delta(H_0)$, which increases with the capability of the quantum devices. 
When the quantum device can prepare all the eigenstates of $H$, $\delta(H_0)$ reaches its maximum value. 
For example, when $H = U_L(\theta)^\dagger P U_L(\theta)$, where $P$ is a Pauli operator, we have $\delta(H_0) = 1$. In this case, our lower bound becomes $\Omega\left(\frac{1}{\epsilon^2}\right)$, given that $\tr{H_0^2}^2 = 4^n$. This lower bound is tight in this case because our parameterized circuit can directly diagonalize $H$.
{In conclusion, a Hamiltonian $ H = U \Lambda U^\dagger $ with a shallow-depth unitary $ U $ is likely to achieve the lower bound in efficiency, making it particularly useful in near-term quantum applications. The 1D nearest-neighbor Heisenberg Hamiltonian, with its sparsely connected terms, is a strong example of this class. Other examples include Hamiltonians with commuting local terms, such as those in the transverse field Ising model, where only nearest neighbors interact, and hence a shallow-depth $ U $ can be constructed. Additionally, matchgate Hamiltonians and certain fermionic Hamiltonians, when decomposed in the Gaussian fermionic basis, permit shallow-depth unitary transformations, which simplify simulation and measurement tasks. Each of these Hamiltonians exemplifies the suitability of our algorithm and further underscores the tightness of the lower bound on circuit depth.}

\section{Conclusion and discussion}
\label{sec:discussion}
We propose a framework for estimating arbitrary linear properties of a quantum state using available quantum circuits. Our framework takes into account both the limitations of quantum devices and the information of the estimated properties. It includes some existing Pauli decomposition protocols~\cite{yen2020measuring,verteletskyi2020measurement,Huang2021Efficient,hillmich2021decision,hadfield2021adaptive,miller2022hardwaretailored,Shlosberg2023Adaptive,wu2023overlapped,gresch2025guaranteed,zhang2023composite,yen2023deterministic}. 

Based on this framework, we propose two classical optimization algorithms that find an optimized decomposition of the observable $H$ with $L$-depth quantum parameterized circuits such that $H \approx \sum_{k=1}^K U_L(\theta_k)^\dagger \Lambda_k U_L(\theta_k)$.
The tensor network approach is well-suited for Hamiltonians that admit compact tensor network representations, while the greedy projection algorithm is designed for general Hamiltonians, particularly in small-sized systems.
We then sample the decomposition terms using the importance sampling technique. Consequently, the expectation value of $\tr{\rho H}$ can be estimated by performing an $L$-depth quantum parameterized circuit sampled from the set $\{U_L(\theta_k)\}_{k=1}^K$, followed by collecting the corresponding diagonal matrix $\Lambda_k$ with computational basis measurements.
Our algorithms can be easily generalized to other circuit structures and non-Hermitian operators. We numerically verify our algorithm by applying it to calculate the ground energy of a given sparse Hermitian and the inner product of a Slater determinant and a pure state.

The performance of our algorithm heavily depends on the decomposition of the target observable. While our greedy decomposition algorithm has been numerically demonstrated to be efficient compared to some conventional methods, it is not guaranteed to be optimal. Therefore, a key future direction is to develop more efficient classical algorithms for optimizing the decomposition process, such as those based on machine learning.
Additionally, our current sample complexity lower bound is loose for low-rank observables, such as fidelity. To address this, we could modify the state discrimination tasks, which form the basis of the proof, to establish a tighter bound. It is also worth exploring the extension of our framework to estimate nonlinear properties and incorporate quantum memories.

In addition to minimizing the sample complexity, we aim to reduce the number of required hardware configurations to estimate a target observable. In practical scenarios, repeating experiments with a fixed configuration is relatively straightforward across various platforms, while changing the configuration often requires additional resources. This goal is equivalent to minimizing the number $ K $ in the decomposition $ H = \sum_{k=1}^K U_k^\dagger \Lambda_k U_k $, where the $ \{U_k\}_k $ are realizable by the hardware. Thus, an important future direction is to explore the relationship between the number of decomposition terms $ K $ and the capability of quantum circuits.

{We note that Ref.~\cite{mangini2024low} proposed an approach for expectation value estimation that combines informationally (over)complete measurements with tensor network representations of both the Hamiltonian and the measurement operators. A notable distinction of our tensor network approach lies in its treatment of measurements: by bypassing the explicit construction of POVM measurements, our method avoids the additional quantum measurement overhead associated with estimating the probabilities of all POVM basis elements.}

% \section*{Methods}
% {
% To support the validity of our approach, we provide proofs for two key propositions that establish the theoretical foundations of our method. Proposition \ref{prop:app_estimation} quantifies the sample complexity required to estimate 
% $\tr{\rho H}$ within a specified accuracy, given an approximate decomposition of $H$. Proposition \ref{prop:app_hermitian} demonstrates the universality of the decomposition approach, showing that any Hermitian matrix $H$ can be approximated by a finite sum of terms involving depth-$L$ unitaries. The proofs of these propositions follow below, with notation consistent with that used in the main text.}

\section*{Acknowledgement}
We would like to acknowledge Zijian Zhang for the valuable discussions regarding the code from their work. 
We would like to thank Xiao Yuan, Zhenyu Du, Jens Eisert, and Hong-Ye Hu for the helpful discussions. We acknowledge the MindSpore Quantum framework as the platform for our numerics.
BW is supported by the National Natural Science Foundation of China (12405014).
ZL and XY acknowledge the support from the National Natural Science Foundation of China Grant No.~12174216 and the Innovation Program for Quantum Science and Technology Grant No.~2021ZD0300804 and No.~2021ZD0300702. FW is supported by BMSTC and ACZSP Grant No.~Z221100002722017.

\appendix

\section{Review of Qubit-Commuting Random Measurements}
\label{app:QC_Review}

Let the Pauli operator expansion of a Hermitian operator $H = \sum_{i=1}^m \alpha_i Q_i$. We use the symbol $P \triangleright Q$ to indicate that $Q$ covers $P$, meaning that each $i$-th qubit of $P$, denoted as $P_i$, is either equal to the identity or to $Q_i$.
We group all $m$ Pauli operators $\cbra{Q_i}$ into overlapped sets $e_1, e_2,\ldots, e_K$, where all Pauli operators in each $e_i$ are commute to each other. There exists a Pauli operator $P^{(i)}$ such that all Pauli operators $Q_j$ in $e_i$ satisfy $Q_j\triangleright P^{(i)}$ for any $i\in [K]$. For instance, when $e=\cbra{Q_1, Q_2}$, $Q_1 = Z_1X_2I_3$ and $Q_2 = Z_1I_2Y_3$, we have $P = Z_1X_2Y_3$.

In the OGM algorithm~\cite{wu2023overlapped} and other grouping measurement algorithms~\cite{hempel2018quantum,izmaylov2019unitary}, sets $e_1, \ldots, e_K$ and measurements $P^{(1)}, \ldots, P^{(K)}$ are explicitly generated with associated probabilities $p_1, \ldots, p_K$. In other random measurement algorithms, while the groups are not explicitly generated, the analysis in Ref. \cite{wu2023overlapped} shows that they implicitly use the same idea. Some algorithms, such as the Derandomized algorithm proposed by Huang et al.~\cite{Huang2021Efficient}, generate the counts of $P^{(1)}, \ldots, P^{(K)}$ when the total number of measurements $T$ is specified, essentially creating a derandomized version of the probabilistic sampling.

Since the Clifford group stabilizes the Pauli group, any Pauli operator can be decomposed into $P = U^{\dagger} \Lambda U$, where $U$ is the tensor product of local Clifford gates. Measuring a quantum state $\rho$ with a Pauli operator $P$ is then equivalent to performing a local Clifford gate $U$, followed by measurement in the computational basis to collect the associated eigenvalues of $P$~\cite{huang2020predicting}. Consequently, qubit-commuting random measurement schemes involve only local Clifford circuits.

\section{Qubit-commuting decomposition}
\label{app:qubit_commut_decomp}
We use the symbol $Q \triangleright P$ to indicate that $P$ covers $Q$, meaning that each $i$-th qubit of $Q$, denoted as $Q_i$, is either equal to the identity or to $P_i$. 
Let the operator $H = \sum_{i=1}^m \alpha_i Q_i$, where $\{Q_i\}_i$ are Pauli operators and $\{\alpha_i\}_i$ are real coefficients. 
Let $e_1, \ldots, e_K$ be (overlapped) qubit-commuting sets such that $e_i \subseteq \{Q_1, \ldots, Q_m\}$. 
We claim that there exists a Pauli operator $P_i$ to cover all elements of the qubit-commuting set $e_i$ for $i \in \{1, \ldots, K\}$.

 The eigendecomposition for $P^{(i)}$ can be represented as $P^{(i)} = U_i^\dagger O_i U_i$ where $U_i$ is a unitary and $O_i$ is diagonal.  For $n$-qubit tensor product unitary $A_j$, denote the $l$-th qubit representation as $A_{jl}$, i.e., $A_j = \otimes_{l=1}^n A_{jl}$. 
Then we have
\begin{align}
H &= \sum_{i = 1}^m \alpha_i Q_i\\
&= \sum_{i=1}^K \sum_{Q_j\in e_i} \alpha_j Q_j\\
&= \sum_{i=1}^K \sum_{Q_j\in e_i} \alpha_j \bigotimes_{l=1}^n Q_{jl}\\
&= \sum_{i=1}^K \sum_{Q_j\in e_i} \alpha_j  U_i^\dagger \pbra{\bigotimes_{l=1}^n D_{jl}} U_i
\label{eq:qc_decomp1}\\
&=\sum_{i=1}^K U_i^\dagger \pbra{\sum_{Q_j\in e_i} \alpha_j D_{j}} U_i\\
&=\sum_{i=1}^K U_i^\dagger \Lambda_i U_i
\label{eq:qc_decomp2}
\end{align}
where $D_{jl}$ equals the single-qubit identity $\Ibb_2$ if $Q_{jl} = \Ibb_2$ and $D_{jl} = O_{jl}$ otherwise,
Eq. \eqref{eq:qc_decomp1} holds since $Q_j\triangleright P^{(i)}$ for any $Q_j\in e_i$, and Eq.~\eqref{eq:qc_decomp2} holds by setting $\Lambda_i = \sum_{Q_j\in e_i} \alpha_j D_j$.

\section{Review of Commuting Measurements}
\label{app:review_commut_meas}
Ref. \cite{yen2020measuring} and Ref. \cite{yen2023deterministic} propose commuting measurement schemes based on grouping or overlapped grouping methods.
In this section, we discuss how to decompose $H=\sum_{j=1}^m \alpha_j Q_j$ into $\sum_{k=1}^M U_k^\dagger \Lambda_k U_k$ where $Q_j$ are Pauli operators, 
$U_k$ are Clifford circuits and $M$ is the number of commuting groups of $\cbra{Q_j}_{j=1}^m$~\cite{yen2020measuring}.

Let $H=\sum_{k=1}^K H_k$ where $H_k=\sum_{Q_j\in g_k} \alpha_j Q_j$ where any two Pauli operators $P_i,P_j$ in set $g_k$ are commute to each other.
To generate the $U_k$ and $\Lambda_k$ for $H_k$, we start by finding sets $\Gcal = \cbra{\vartheta_1,\ldots, \vartheta_n}$ and $\Qcal = \cbra{\sigma_1,\ldots, \sigma_n}$ where $\vartheta_i$ are Pauli operators and $\sigma_j$ are single-qubit Pauli operators such that
\begin{itemize}
    \item $[\vartheta_i, \vartheta_j] = 0$ for any $i,j\in [n]$;
    \item $ [\vartheta_i, Q_l]=0$ for any $i\in[n], Q_l\in g_k$; 
    \item $\cbra{\vartheta_i,\sigma_j} = 0$ for any $i\ne j\in [n]$;
    \item $[\vartheta_i,\sigma_i] = 0$ for any $i\in [n]$;
    \item $[\sigma_i,\sigma_j] = 0$ for any $i,j\in [n]$.
\end{itemize}
According to Ref.~\cite{bravyi2017tapering}, we can give the set $\Gcal$, which serves as the generator set for Pauli operators in $g_k$, meaning that all $Q_j$ can be expressed as a product of some elements $\vartheta_i$ in $\Gcal$. It can also be easily proved that such set $\Qcal$ also exists. 
Let $U_k'=\prod_{l=1}^n \frac{1}{\sqrt{2}}\pbra{g_l + \sigma_l}$.
Suppose $\sigma_l = C_l^\dagger D_l C_l$. Let
$L_k = \prod_{l=1}^n C_l$ and $U_k = L_kU_k'$. Then we have
$U_k' g_l U_k'^\dagger = \sigma_l$, hence $U_k g_l U_k^\dagger = L_k \sigma_l L_k^\dagger = D_l$, and
\begin{align*}
    U_k H_k U_k^\dagger = \sum_{Q_j \in g_k} \alpha_j \prod_{\vartheta_l\in S_j} D_l,
\end{align*}
where $Q_j = \prod_{\vartheta_l\in S_j} g_l$. Such $S_j$ exists by the definition of $\cbra{\vartheta_i}_i$.
Let $\Lambda_k = \sum_{Q_j \in g_k} \alpha_j \prod_{\vartheta_l\in S} D_l$, then $H = \sum_{k=1}^K U_k^\dagger \Lambda_k U_k$.

\section{Proof of Proposition~\ref{proposition:SampleComplexity_LB}}
\label{app:LowerBound}

\begin{definition}
For probability distributions $p=\{p_1,\cdots,p_N\}$ and $q=\{q_1,\cdots,q_N\}$, we define the total variation between them
 by $\operatorname{TV}(p,q)=\frac{1}{2}\sum_{i=1}^N\abs{p_i-q_i}$.
\end{definition}

\begin{lemma}[Le Cam's two-point method, Lemma 1 in \cite{yu1997assouad} and Lemma 5.3 in \cite{Chen2021exponential}] 
For a learning algorithm without quantum memory being described as a rooted tree $\Tcal$,
let $p_{\rho}$ be a probability distribution on the leaves, such that $p_{\rho_x}(l)$ equals the probability of arriving at the leaf $l$.
The probability that the learning algorithm that can distinguish quantum state sets $\cbra{\rho_x}_x$ and $\frac{\Ibb}{2^n}$ correctly is upper bounded by $\frac{1}{2} + \frac{1}{2}\TV(\Ebb_x \sbra{p_{\rho_x}}, p_{\Ibb/2^n})$.
\label{lem:LeCamTwo_point}
\end{lemma}

We derive our result by extending the proof of Theorem 5.5 from Ref. \cite{Chen2021exponential}. The central idea is to show that for a Hamiltonian 
 $H$, there exist two sets of quantum states $\cbra{\rho_x}_x$ and $\cbra{\frac{\Ibb}{2^n}}$, such that distinguishing between the expectations $\tr{\rho_x H}$ or $\tr{\frac{H}{2^n}}$ requires at least $T$ samplings. This task is performed under the condition that $L$-depth quantum circuits are applied before measurement in the computational basis. By Lemma \ref{lem:LeCamTwo_point}, the task of quantifying how distinguishable the distributions generated by the quantum algorithm using \( T \) copies of the quantum states is can be simplified to estimating the total variation distance \( \text{TV}(\mathbb{E}_x \left[p_{\rho_x}\right], p_{\mathbb{I}/2^n}) \) between two distributions, which provides the lower bound of sample complexity $T\geq \Omega\pbra{\frac{1}{a^2\delta(H)}}$ for the observable measurement task. We will give details later on how to generate this lower bound in the proof part.

We also need to relate the parameter \( a \) to the estimation error \( \epsilon \).
Without loss of generality, assume $\tr{H}=0$ for the Hamiltonian under consideration. Denote $H=H_1=-H_2$. We now define two state sets:
\begin{itemize}
\item Set 1: $\{\mathbb{I}/2^n\}$,
\item Set 2: $\{\rho_x\}_{x=\cbra{1,2}}$, each with equal probability,
\end{itemize}
where $\rho_x = \left(\mathbb{I}+3xH_1\right)/2^n$, and $a\le\frac{1}{3\vabs{H}_2}$ to ensure that the matrices in Set 2 are valid density matrices, where $\vabs{H}_2$ denotes the spectral norm (two-norm) of $H$.

Consider the state discrimination task involving these two sets. A referee randomly selects one of the sets, and then chooses a state $\rho$ from the selected set according to the corresponding distribution. 
The referee provides $T$ copies of $\rho$,
and our task is to determine which set $\rho$ comes from, using these $T$ copies. The quantum device available for this task is a $n$-qubit quantum device of limited capability, whose all possible output states when acting on $\ket{0^n}$ form a set $\mathcal{S}$.

Note that $\tr{H\frac{\mathbb{I}}{2^n}}=0$, $\tr{H\frac{\mathbb{I}+3aH_1}{2^n}}=\frac{3a\tr{H^2}}{2^n}$ and $\tr{H\frac{\mathbb{I}+3aH_2}{2^n}}=-\frac{3a\tr{H^2}}{2^n}$. One possible way to accomplish the state discrimination task is to measure $H$ to precision $\epsilon=\frac{a\tr{H^2}}{2^n}$ and check whether the approximate expectation value lies in $\big[-\frac{a\tr{H^2}}{2^n},\frac{a\tr{H^2}}{2^n}\big]$. If it holds, we assign $\rho$ to Set 1, otherwise we assign $\rho$ to Set 2. 

Thus if we can use this quantum device to measure $\tr{H\rho}$ with an accuracy of $\epsilon=\frac{a\tr{H^2}}{2^n}$ and a success probability of at least $2/3$ for any given $\rho$, then we can accomplish the state discrimination task with the same success probability of at least $2/3$ using this quantum device.

In the following, we show the sample complexity $T$ for accomplishing the state discrimination task by this quantum device with a success probability of at least $2/3$, is lower bounded by $T\ge\Omega\left(\frac{1}{a^2\delta(H)}\right) = \Omega\left(\frac{\tr{H^2}^2/4^n}{\epsilon^2\delta(H)}\right)$.
Consequently, the sample complexity for observable measurement is also lower-bounded by this quantity.

\begin{proof}
[Proof of Proposition 3 in the main text]
When the quantum device is capable of performing arbitrary POVM measurements, any state discrimination protocol can be represented by a decision tree, as shown in Figure 4 of \cite{liu2024separation}. However, since our quantum device is limited to performing parameterized unitary operations $U_L(\theta)$ followed by measurements in the computational basis, the experiment at each node of the decision tree is modified accordingly. Specifically, instead of performing an arbitrary POVM, the measurement at each node is represented by the set of outcomes $\{\alpha_i(\theta)\}_{i=1}^{2^n}$, where $\alpha_i(\theta)$ is the $i$-th column of $U_L(\theta)^\dagger$.

Given a state discrimination protocol using this quantum device, and the $\rho$  selected by the referee, we label the leaves of the decision tree by $l$. 
By Lemma \ref{lem:LeCamTwo_point}, to successfully accomplish the state discrimination task with a probability of at least $2/3$, we need to ensure the depth $T$ of the decision tree (i.e., the sample complexity) is large enough to satisfy the condition in
\begin{equation}\label{eq:total_variation}
\operatorname{TV}\left(\mathbb{E}_x\sbra{p_{\rho_x}},p_{\mathbb{I}/2^n}\right)\ge\frac{1}{3},
\end{equation}
where $\rho_x = \frac{\Ibb + 3aH_x}{2^n}$ for $x\in\cbra{1,2}$ denotes the quantum state in set 2.
For each leaf $l$ in the decision tree, there is a unique path from the root node $n_0$ to the leaf node $n_T=l$. 
Let the sequence of nodes along this path be denoted as $n_0,\cdots,n_T$ and the edges between them as $e_1,\cdots,e_T$. 
For each $t=1,\cdots, T$, we associate the projector on edge $e_{t}$ by $\ketbra{\alpha_{e_t}(\theta_{n_{t-1}})}{\alpha_{e_t}(\theta_{n_{t-1}})}$.
To give a upper bound of the total variation TV$(\Ebb_x[p_{\rho_x}], p_{\Ibb/2^n})$, we proceed by calculating the following quantity:
\begin{equation}
\begin{aligned}
\frac{\mathbb{E}_x\sbra{p_{\rho_x}(l)}}{p_{\mathbb{I}/2^n}(l)}&=\mathbb{E}_x\prod_{t=1}^T\left(\frac{1/2^n+1/2^n3a\bra{\alpha_{e_t}(\theta_{n_{t-1}})}H_x\ket{\alpha_{e_t}(\theta_{n_{t-1}})}}{1/2^n}\right)\\
&=\mathbb{E}_x\prod_{t=1}^T\Big(1+3a\bra{\alpha_{e_t}(\theta_{n_{t-1}})}H_x\ket{\alpha_{e_t}(\theta_{n_{t-1}})}\Big)\\
&=\mathbb{E}_x\exp\left[\sum_{t=1}^T\log\Big(1+3a\bra{\alpha_{e_t}(\theta_{n_{t-1}})}H_x\ket{\alpha_{e_t}(\theta_{n_{t-1}})}\Big)\right]\\
&\geq\exp\left[\sum_{t=1}^T\mathbb{E}_x\log\Big(1+3a\bra{\alpha_{e_t}(\theta_{n_{t-1}})}H_x\ket{\alpha_{e_t}(\theta_{n_{t-1}})}\Big)\right]\\
&=\exp\left[\sum_{t=1}^T\frac{1}{2}\log\Big(1-9a^2\bra{\alpha_{e_t}(\theta_{n_{t-1}})}H\ket{\alpha_{e_t}(\theta_{n_{t-1}})}^2\Big)\right]\\
&\geq\exp\left[-9a^2\sum_{t=1}^T\bra{\alpha_{e_t}(\theta_{n_{t-1}})}H\ket{\alpha_{e_t}(\theta_{n_{t-1}})}^2\right]\\
&\geq\exp\left[-9a^2T\delta(H)\right]\\
&\geq1-9Ta^2\delta(H)\\
\end{aligned}
\end{equation}
where the second inequality holds since $\log(1-u)\geq-2u$ for all $u\in[0,0.79]$.
Using this, we can estimate the total variation as:
\begin{equation}\label{eq:TV_Epsi_I}
\begin{aligned}
\operatorname{TV}\left(\underset{x}{\mathbb{E}}\sbra{p_{\rho_x}},p_{\mathbb{I}/2^n}\right)=\sum_{l}p_{\mathbb{I}/2^n}(l)\max\left\{0,1-\frac{\underset{x}{\mathbb{E}}\sbra{p_{\rho_x}(l)}}{p_{\mathbb{I}/2^n}(l)}\right\}\leq9Ta^2\delta(H).
\end{aligned}
\end{equation}
Combined with Eq. \eqref{eq:total_variation}, we need to ensure
\begin{equation}
\frac{1}{3}\le\operatorname{TV}\left(\underset{x}{\mathbb{E}}\sbra{p_{\rho_x}},p_{\mathbb{I}/2^n}\right)\le9Ta^2\delta(H),
\end{equation}
which gives the bound
\begin{equation}
T\ge\Omega\left(\frac{1}{a^2\delta(H)}\right)=\Omega\left(\frac{\tr{H^2}^2/4^n}{\epsilon^2\delta(H)}\right).
\end{equation}
\end{proof}

\section{Preliminary to inner product calculation}
\label{app:pre_inputstate}
Given $\ket{\psi}:=U_{\psi}\ket{0}$ and an ancilla initialized to $\ket{0}_a$, by performing Hadamard gate on $\ket{0}_a$, followed by Control-$U_{\psi}$ gate, we obtain
$\ket{\psi'}=\frac{\ket{0^{n}} + \ket{1}\ket{\psi}
}{\sqrt{2}}$.

Since the observable $H$ is not Hermitian and $\tr{\rho H}$ cannot be directly approximated with existing random measurement algorithms. When executing the existing algorithm to calculate the inner product, we decompose the observable $H$ into two Hermitians $H_1:= (H + H^\dagger)/2$ and $H_2 := -\im (H - H^\dagger)/2$, then $H = H_1+\im H_2$, and it can then be approximated in two parts. The total error $\varepsilon = \abs{\hat{v} -\tr{\rho H}}=\sqrt{\pbra{\hat{v}_1 - \tr{\rho H_1}}^2+\pbra{\hat{v}_2 -\tr{\rho H_2}}^2}$ where $\hat{v}_j$ is the estimation of $\tr{\rho H_j}$ for $j\in\cbra{1,2}$.

\section{Supplementary numerics}
\label{app:supp_numerics}

\begin{figure}[t]
    \centering
 \includegraphics[width = 1.0\textwidth]{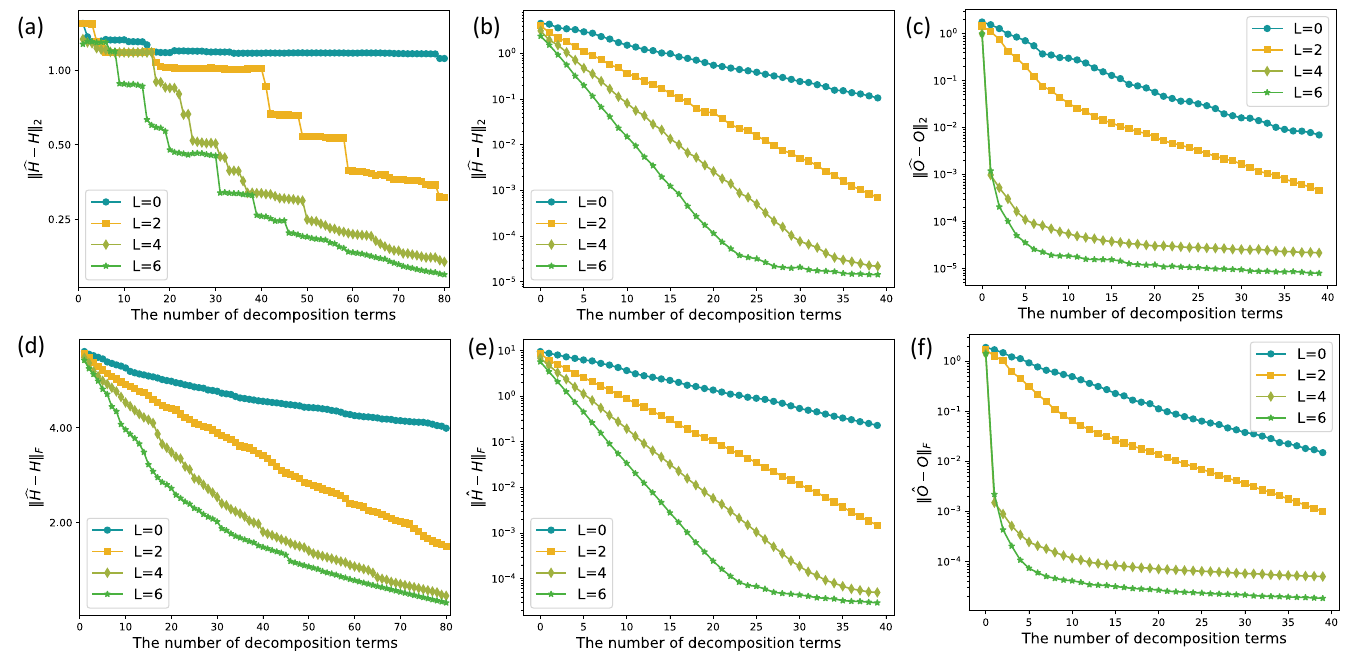}
    \caption{{The variations in Frobenius distance as the number of decomposition terms increases for (a) a sparse Hermitian $H$ and (b) a random generated observable, and (c) a non-Hermitian operator $O$ associated with a Slater determinant.}}
    \label{supp_fig:Fnormdistance}
\end{figure}

Here, we present the distances based on $2$-norm and $F$-norm of $H^{(k)}$ and $H$ as $k$ increases for sparse Hamiltonian (a),(d), dense Hamiltonian (b),(e) and operator associated with Slater-determinant (c), (f). As illustrated in Fig.~\ref{supp_fig:Fnormdistance}, the Frobenius distance between $H^{(k)}$ and $H$ decays exponentially.

\begin{figure}
    \centering
    \includegraphics[width=0.6\linewidth]{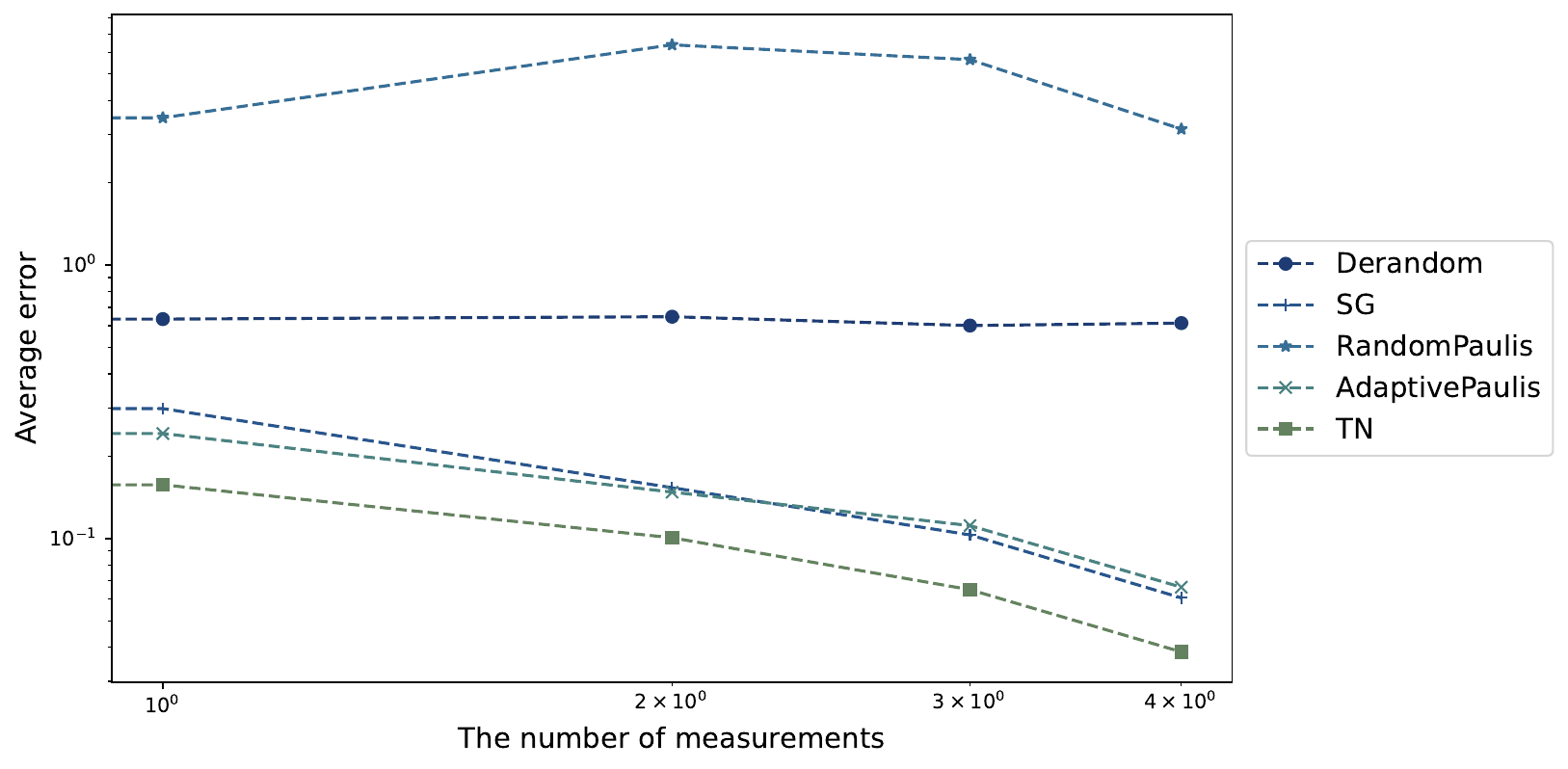}
    \caption{{Comparison of the TN algorithm with Derandom~\cite{Huang2021Efficient}, SG~\cite{gresch2025guaranteed}, RandomPauli~\cite{huang2020predicting}, and AdaptivePaulis~\cite{Shlosberg2023Adaptive} in terms of the average estimation error as a function of the number of measurements. The evaluation is performed over five experiments, each using a different randomly generated quantum input state. The system has 8 qubits. The TN algorithm uses a quantum parameterized circuit with depth $L=1$ and employs $K = 3$ decomposition terms.}}
    \label{fig:average_tn_numerics}
\end{figure}

Here we give the performance of the tensor-network algorithm and other existing algorithms in the average of five group experiments, concerning the random input states, for the same Hamiltonian $W$ defined in the main text. The comparison with existing qubit-wise commuting algorithms is illustrated in Fig.~\ref{fig:average_tn_numerics}.

\begin{figure}
    \centering
    \includegraphics[width=1.0\linewidth]{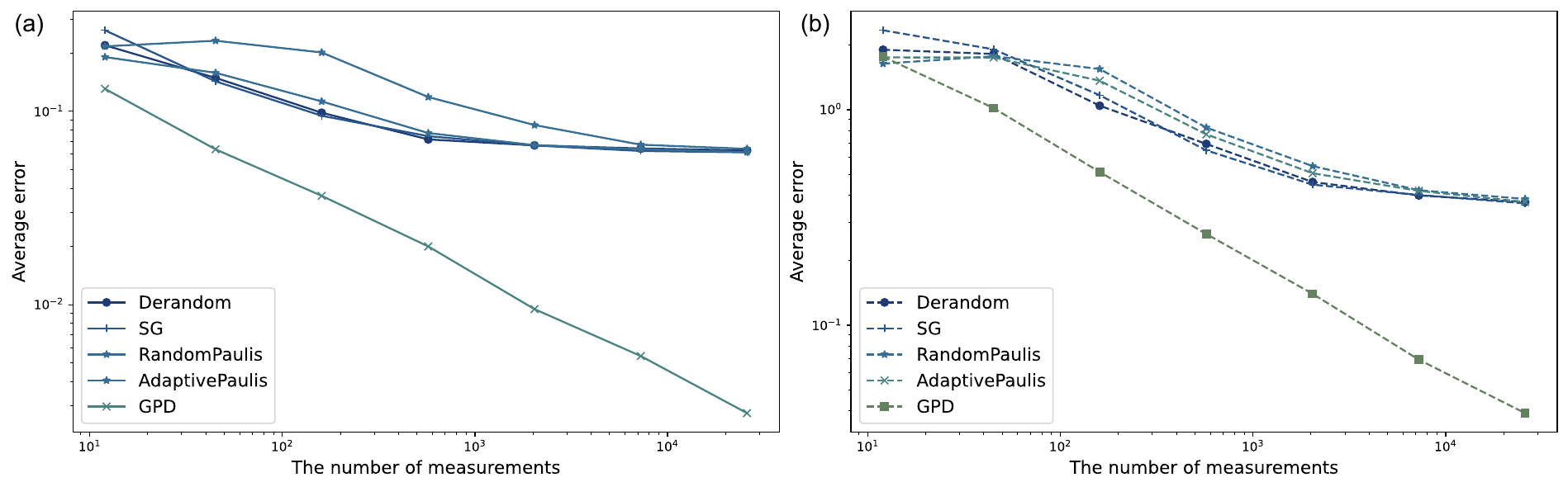}
    \caption{{Comparison of the GPD algorithm with Derandom~\cite{Huang2021Efficient}, SG~\cite{gresch2025guaranteed}, RandomPauli~\cite{huang2020predicting}, and AdaptivePaulis~\cite{Shlosberg2023Adaptive} in terms of average estimation error versus the number of measurements, evaluated over five randomly generated datasets. (a) Random sparse Hamiltonians and (b) Random dense (general) Hamiltonians. The system has 4 qubits. The GPD algorithm uses a quantum parameterized circuit with depth $L = 4$ and employs $K = 20$ decomposition terms.}}
    \label{fig:average_error_5data}
\end{figure}

To further demonstrate the average-case performance of greedy projection Decomposition (GPD) algorithms, we present the average estimation errors over five randomly generated Hamiltonians, including both sparse and dense cases. The number of qubits is fixed at $n=4$. For each Hamiltonian, we construct a $2^n \times 2^n$ matrix $A$. In the sparse case, we randomly select $n$ non-zero elements from the interval $[0,1]$ to populate $A$, while in the dense case, all elements of $A$ are independently sampled from $[0,1]$. To ensure Hermiticity, we define the Hamiltonian as $H = (A + A^\dagger)/2$.
The input state is the same as that used for the random dense Hamiltonian in the main text and is also randomly generated. The GPD algorithm uses a quantum parameterized circuit with depth $L = 4$ and employs $K = 20$ decomposition terms. 
The comparison in Fig.~\ref{fig:average_error_5data} demonstrates that, on average, our algorithm significantly outperforms existing qubit-wise commuting methods. 
As most of these methods exhibit similar performance, we include only a few representative algorithms in the figure to enhance clarity and readability.

\begin{figure}
    \centering
    \includegraphics[width=0.8\linewidth]{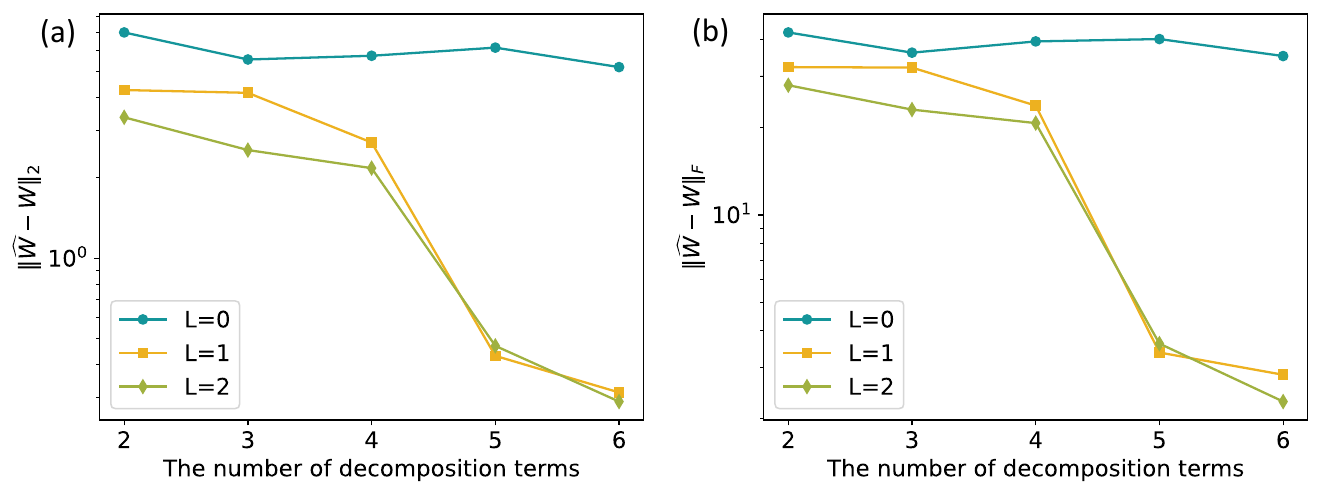}
    \caption{{The variations in Frobenius distance for the estimated Hamiltonian $\hat{W}$ and the ideal low bond dimension Hamiltonian $W$ as the number of decomposition terms increases with the TN method.}}
    \label{fig:TN_numerics}
\end{figure}

Fig. \ref{fig:TN_numerics}(a-b) illustrates the approximation error of the Hamiltonian $W$ as a function of the number of decomposition terms, using ansatz circuits of depths 0, 1, and 2, which shows the estimation error for the Hamiltonian declines with the decomposition terms, and one layer of quantum parameterized circuit is enough for the estimation. 

%\bibliographystyle{unsrt}
%\bibliography{ref}

\end{document}